\documentclass[conference]{IEEEtran}
\IEEEoverridecommandlockouts
\usepackage{cite}
\usepackage{amsmath,amssymb,amsfonts}
\usepackage{algorithmic}
\usepackage{graphicx}
\usepackage{textcomp}
\usepackage{xcolor,multirow}

\def\BibTeX{{\rm B\kern-.05em{\sc i\kern-.025em b}\kern-.08em
    T\kern-.1667em\lower.7ex\hbox{E}\kern-.125emX}}
\begin{document}

\title{BalanceDN: Load-Balancing Allocation of Interest for Fast Discovery in Content Centric Networks\\
\thanks{This paper was peer-reviewed, accepted, and presented at IEEE Word Forum of Internet of Things 2023. \\Cite as: M. Gunti and R. Rojas-Cessa, "BalanceDN: Load-balancing Allocation of Interest for Fast Discovery in Content Centric Networks," in IEEE Word Forum of Internet of Things, 5 pp., Oct. 12-27, 2023, Alveiro, Portugal.}
}

\author{\IEEEauthorblockN{Murali Gunti}
\IEEEauthorblockA{\textit{Networking Research Laboratory} \\
Department of Electrical and Computer Engineering \\
New Jersey Institute of Technology\\
Newark, NJ 07102 \\
mg496@njit.edu
}
\and
\IEEEauthorblockN{Roberto Rojas-Cessa}
\IEEEauthorblockA{\textit{Networking Research Laboratory} \\
Department of Electrical and Computer Engineering\\
New Jersey Institute of Technology\\
Newark, NJ 07102 \\
rojas@njit.edu}
}

\maketitle

\begin{abstract}
In Named Data Networking (NDN), data is identified by unique names instead of IP addresses, and routers use the names of the content to forward Interest packets towards the producers of the requested content. However, the current content search mechanism in NDN is complex and slow. This mechanism not only creates congestion but also hinders practical deployment due to its slowness and cumbersome nature. To address this issue, we propose a methodology, called BalanceDN, that distributes content through the network such that sought content can be found quickly. BalanceDN uses a distributed allocation of resolvers as those used by the domain name system but differs in how content is distributed. Our approach avoids flooding the network with pending interest requests and also eliminates the need for blind search when the location of content is unknown. We tested our approach on a simulation platform for NDN called ndnSIM. The results show that the proposed routing scheme utilizes far fewer network resources compared to the NDN network when retrieving content. The proposed scheme accomplishes this performance gain by leveraging a load-balanced hashing mechanism to distribute and locate the name of the content on the distributed nameserver lookup service nodes.

\end{abstract}


\section{Introduction}
\label{sec:introduction}

Named Data Networking (NDN) is proposed as a future Internet architecture that will replace the current Network Layer (IP) \cite{NDN}. The current IP-based model focuses on sending data packets to specific IP addresses or endpoints, while NDN focuses on the data itself, allowing the data to be routed by its own name, regardless of the endpoint \cite{NDNSurvey}. This means users can access data by simply specifying its name, without needing to know the specific endpoint or IP address where it is stored \cite{ndn3}.

Communication in the NDN architecture relies on the exchange of two packet types: Interest and Data \cite{icn2,ccn1,ccn3}. When a user wants to request a particular piece of data, they create an Interest packet with the data name and transmit it to the network. NDN nodes use this name to forward the Interest toward the data producer \cite{IntroNDN}. Once the Interest reaches a node that has the requested data, the node produces a Data packet that includes the name, content, and a signature. This Data packet is then routed back on the same path to the user \cite{ndngen1}.

To handle the forwarding of Interest and Data packets, each NDN node is equipped with three data structures: a Pending Interest Table (PIT), a Forwarding Strategy module, which includes the Forwarding Information Base (FIB), and a Content Store (CS) \cite{ndn7}. These structures help the NDN node determine when and where to forward each Interest packet. The Pending Interest Table (PIT) tracks Interest packets that are awaiting a response from the network.

The Forwarding Strategy module contains all the routing protocols and the FIB to make decisions on when and where to route the Interest packet. The Forwarding Information Base (FIB) is a routing table that maintains a copy of the forwarding information. The Content Store (CS) stores and caches data that has been previously requested by the user \cite{ndn6}.

Routing in NDN \cite{RoleRouting} is based on the unique names assigned to data packets, which allows routers to forward Interest packets towards the data producers responsible for providing the requested data \cite{ndn8}. Essentially, the network will flood the Interest until the producer is found, and then the Data packet will be routed back to satisfy the request. This can create congestion and tax the capacity of the network, especially if the data is not cached on any of the nodes. Data is requested by sending an Interest packet expressing interest in a specific piece of data. The Interest packet contains a name that identifies the requested data, and when a node receives it, it checks its content store for the requested data \cite{ndn8}. If found, it is returned to the data consumer along the same path that the Interest packet took, forming a reverse path.

To handle the forwarding of Interest and Data packets, each NDN node is equipped with three data structures: a Pending Interest Table (PIT), a Forwarding Strategy module, which includes the Forwarding Information Base (FIB), and a Content Store (CS). These structures help the NDN node determine when and where to forward each Interest packet. The Pending Interest Table (PIT) tracks Interest packets that are awaiting a response from the network \cite{ndn11}. 

The Forwarding Strategy module makes decisions on the routing, and the FIB makes decisions on when and where to route the Interest packet. The Forwarding Information Base (FIB) is a routing table that maintains a copy of the forwarding information \cite{RoleRouting}. The Content Store (CS) stores and caches data that has been previously requested by the user \cite{ndn12}.

The current content search mechanism in NDN is complex and slow, critically hindering practical deployment. To address this issue, this paper proposes a methodology to place and locate content within the NDN principles, which utilizes a load-balanced hashing mechanism to distribute and locate content names on distributed nameserver lookup service nodes. We call this method BalanceDN. This mechanism builds on the proposed work on Named Data Networking Service (NDNS) \cite{NDNS}. BalanceDN results in an efficient routing scheme that reduces congestion and network resource usage. Our extensive evaluations on three different network topologies and various content distributions show that BalanceDN is effective and efficient.

The contribution of this paper is the proposal of BalanceDN, a framework that a) distributes content in a NDN network for simple location and uniform and quick access by the average user, and b) a methodology to back-trace requests to locate content while keeping the design principles of NDN, and c) the testing of the access time in three different network topologies with various content distributions. The results show that the proposed framework achieves significant performance improvement as compared to the conventional content search method.

The remainder of this paper is organized as follows. Section \ref{sec:related-work} describes related work. Section \ref{sec:proposed-scheme} introduces the proposed load-balanced hashing mechanism used to place and locate content in NDN. Section \ref{sec:results} presents our evaluation results. Section \ref{sec:conclusions} presents our conclusions.

\section{Related Work}
\label{sec:related-work}

Named Data Networking Service (NDNS) is a distributed name resolution service designed for the Named Data Networking (NDN) architecture \cite{NDNS}. NDNS is a DNS-like system for its application in NDN. The motivation for proposing an NDN-based DNS is to provide a scalable routing table for NDN because the FIB tables on NDN nodes are limited. It also aims to provide mobility support for data producers and persistent storage for cryptography keys. In NDNS, the top-level NDNS servers provide hints about where the content can be found without knowing for sure if the content exists or not, which still relies on flooding to find content. Thus, this paper does not address the flooding problem in NDN. The current investigation, however, particularly aims to solve the problem of flooding by allocating multiple DNS-like NDN resolvers to directly find the content. 

\section{Proposed Scheme for Finding Interest Content}
\label{sec:proposed-scheme}

\begin{figure}[htbp]
    \centering
    \includegraphics[width =0.99 \columnwidth]{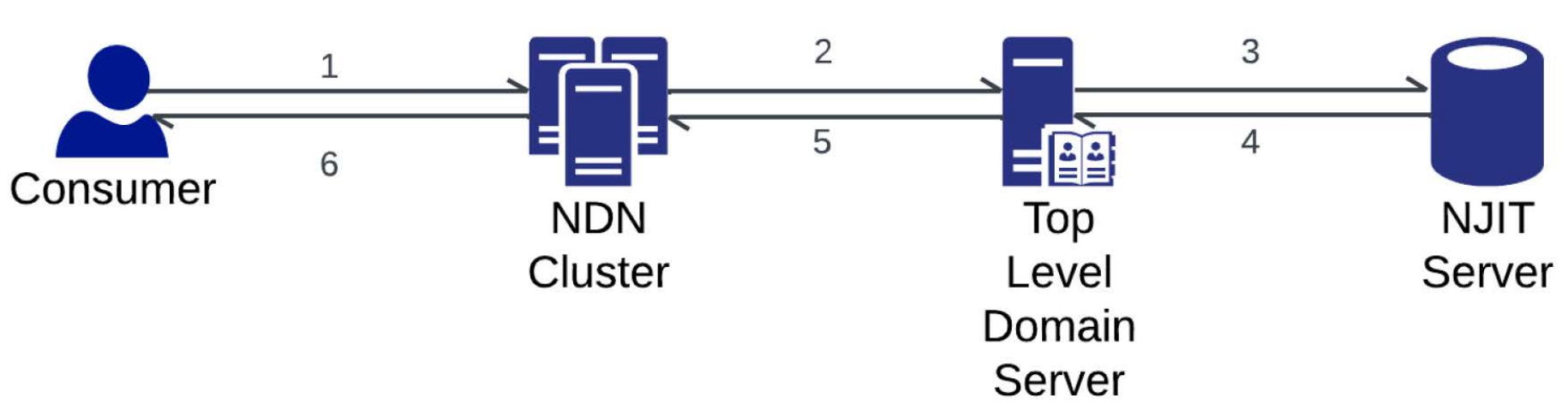}
    \caption{Example of the flow of uploading data through the proposed NDN routing scheme.}
    \label{fig:model}
\end{figure}

\begin{figure}[htbp]
    \centering
    \includegraphics[width =0.99 \columnwidth]{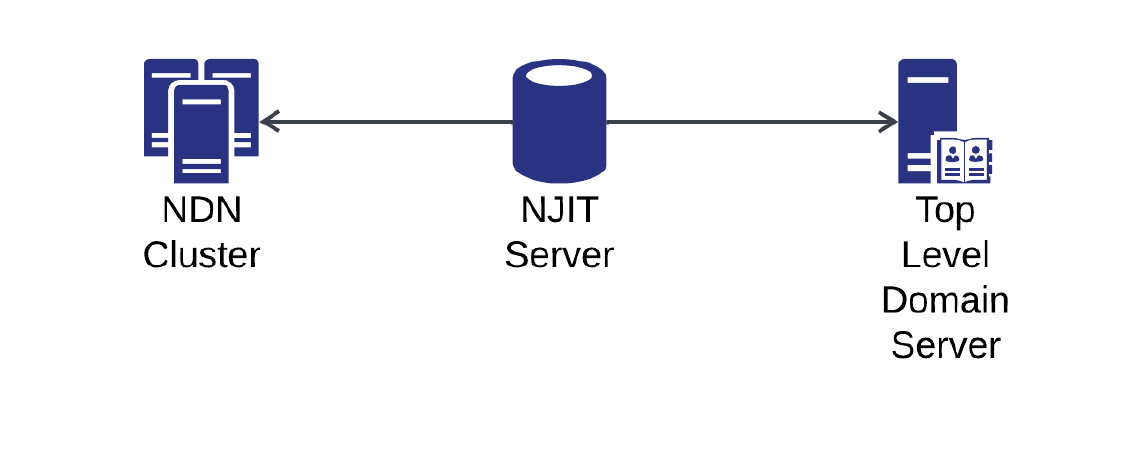}
    \caption{Architecture of the host-resolver.}
    \label{fig:resolver}
\end{figure}

\begin{figure}[htbp]
    \centering
    \includegraphics[width =0.89 \columnwidth]{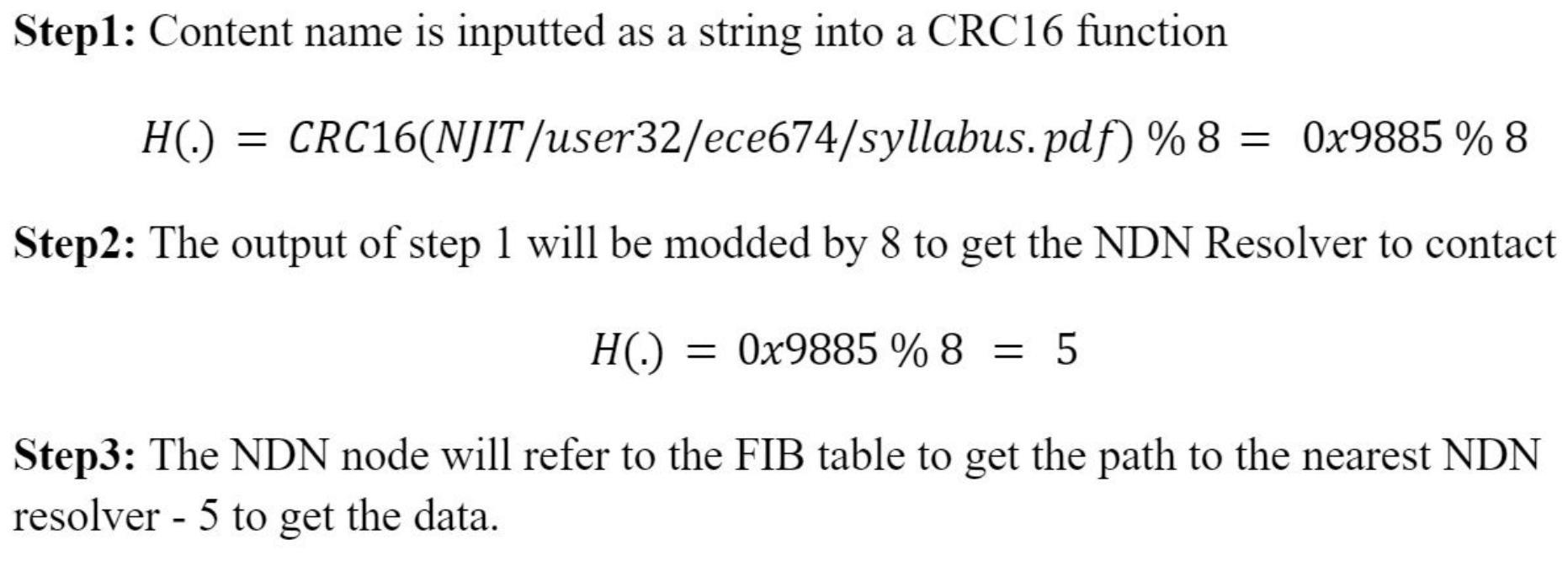}
    \caption{Pseudocode of the content discovery mechanism.}
    \label{fig:algorithm}
\end{figure}

The implementation of the proposed DNS-like service for NDN involves adding new servers and protocols to the current NDN network. A cluster of 8 NDN servers is introduced, which functions as NDN resolvers. Similar to DNS resolvers, NDN resolvers maintain a table pairing the interfaces of Top-Level servers, name servers, other NDN resolvers, and producers who have the content data to a content name. The flow of data through the network is shown in Figure \ref{fig:model}, and each step is indicated by a number in the order it occurs. Step 1 is when an Interest is sent to an NDN cluster.
At this point, the hashing algorithm processes the user request and outputs a hash number, which is associated with one of the eight NDN resolvers. Then the Interest is routed to the appropriate NDN resolver. The resolver searches its FIB tables and communicate with the appropriate Top-Level Domain (TLD) Server to identify the Name Server on record for the requested content, as shown in Step 2. In Step 3, the TLD Server looks through its table and route to the appropriate Name Server that provides the location of the content. Steps 4, 5, and 6 show the flow of the data back to the consumer as the data gets cached at each of the servers for faster lookup next time. If the NDN Resolver has the interface to the Name Server, other NDN resolver, or the producer, then it skips contacting the TLD Server and fetches the data directly from the nearest producer, reducing the network overhead. This method of routing in NDN ensures that the network always has the information about the content in the network. Therefore, when a producer aims at sharing content, the NDN node on the producer's end performs the load-balanced hashing algorithm and outputs the TLD server and NDN cluster to establish a path to the newly uploaded content. This self-sustaining model ensures that content can be found directly, with minimal additional searches. This process is shown in Figure \ref{fig:resolver}, where a producer contacts the appropriate NDN cluster and TLD Server to establish a path to data.
The load-balanced hashing algorithm used for this model is displayed below. The routing algorithm helps distribute content evenly across the eight NDN resolvers so that the NDN cluster can use the same algorithm to hash the content and match the NDN resolver that has the content of interest. In this way, the algorithm distributes content evenly and also helps route to data quickly as the location is known for direct access. This process avoids inquiring every NDN node, as in the conventional scheme.
\begin{equation}
 H(.)=CRC16(s) \mod N   
\end{equation}
where $N$ is the number of resolvers in the system and CRC16 is the cyclic redundancy check (CRC) for 16 bits.
The routing algorithm uses CRC16 to generate a hash, which then is modded by N, the number of different types of NDN resolvers, to get the server to contact. CRC16 is used because it is a balanced method for distributing the content over the network. This approach ensures that the network is balanced in terms of capacity and no particular NDN node is under or over utilized. Below is an example of how the hashing methodology works, as shown in Figure \ref{fig:algorithm}.

The NDN Clusters can be distributed in various physical locations depending on factors such as the distribution of users, the need for redundancy and backup, and the need for additional storage capacity. Network operators can monitor the network and distribute the NDN resolvers as needed to meet the Quality of Service (QoS) required for the network.

\section{Evaluation Results}
\label{sec:results}

To evaluate the effectiveness of the proposed model, the network utilization of the NDN network was assessed by measuring the bandwidth used by both Interest and Data packets as they travel through the network. This was done by calculating the number of hops needed to locate content, as it indicates the resources utilized by the Interest packets while searching for the content.
In the first scenario, we implemented the routing scheme on NSFnet by adding a few users to each of the root nodes. We chose this topology because it has been used in many other NDN papers and is an established topology, ensuring unbiased results. We conducted a series of tests on this small network with three primary objectives: first, to verify the effectiveness of the proposed method, secondly, to demonstrate the limitations of traditional NDN routing, and finally, to present a working implementation of the idea.

In the second scenario, we conducted a more challenging test using the same NSFnet topology. Each node requested a unique content from every producer in the network, including from those in its own subnet. This test allows us to measure the number of hops taken to retrieve the content for both the proposed routing scheme and the built-in method, as the content had to travel across multiple subnets. This test also provides insights into how much of the network each routing scheme utilized to retrieve content.

In the third scenario, the proposed routing scheme was evaluated on the OTEGlobe topology, which is a much larger topology than the previous one. The test assesses the scalability of the proposed method and helps to ensure that the results obtained in the previous scenario hold up when applied to a larger network. To achieve this, the same test as in Scenario 2 was conducted.
In the fourth scenario, we tested the efficiency and performance of the hashing algorithm when the data is not distributed uniformly. 

\begin{figure}[htbp]
    \centering
    \includegraphics[width =0.69 \columnwidth]{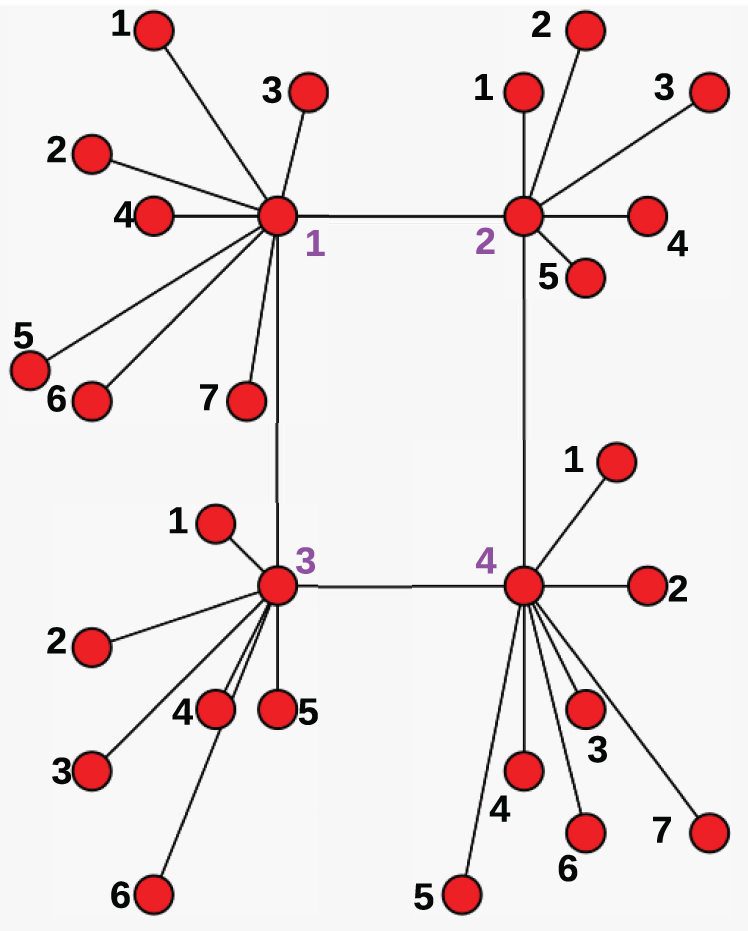}
    \caption{Simple test network with 3 hops as the largest distance.}
    \label{fig:testnet}
\end{figure}

\begin{figure}[htbp]
    \centering
    \includegraphics[width =0.79 \columnwidth]{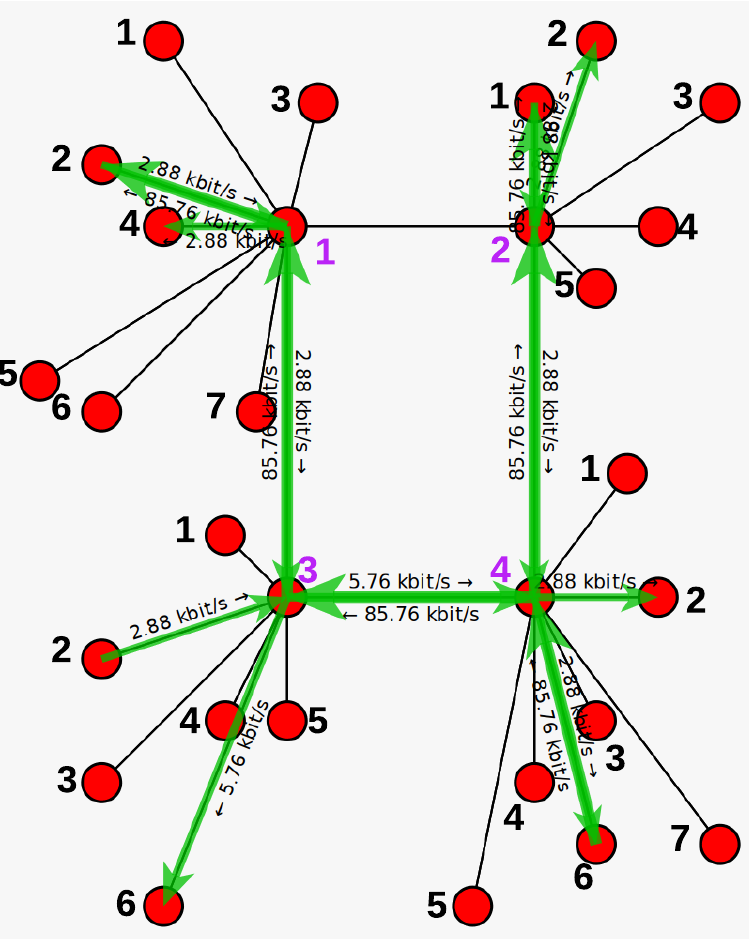}
    \caption{Example of used discovery routes (green color) on the test network.}
    \label{fig:testnet-green}
\end{figure}

Figure \ref{fig:testnet} portrays the constructed test network, where each red dot represents an NDN node, and the block lines depict the interconnecting links between them.
Once the network was constructed, several simulations were conducted to gain insight into the workings of the simulator and the structure of the code, as depicted in Figure \ref{fig:testnet-green}. The green lines highlight the route taken by the Interest packets to locate the data. The thicker green lines signify the path of both Interest and Data packets, whereas the thinner green line illustrates the path of the Interest alone. Additionally, the simulator portrays the data bandwidth throughout the network.

\subsection{Scenario 1: Efficiency in a Moderate Size Network}
\label{subsec:sce1}

The simulation parameters were established and a topology
resembling the NSFnet was built in ndnSIM, as depicted in Figure \ref{sim1-network}. This topology was designed with 11 router nodes represented by yellow dots, 22 consumer nodes depicted in red, and 11 NDN clusters illustrated in teal, amounting to a total of 54 nodes in the network.

In this experiment, the efficiency of the proposed scheme was evaluated by initiating a consumer request for an Interest from the producer of size 1024 bits. To assess the performance of the scheme, three experiments were conducted, each with increasing distance between the consumer and the producer.
The first experiment involved a consumer request that was only one hop away from the producer, the second experiment was performed with the consumer two hops away from the producer, and the third experiment involved the producer being located four or more hops away. The purpose of increasing the number of hops between the consumer and the producer was to analyze the number of hops the Interest took as the distance between them increased. To evaluate the results, the top 5\% of the hops were averaged and rounded up. The performance of the proposed scheme was compared against the built-in scheme to assess its effectiveness.

In Scenario-1-near, the producer is represented by the purple color and the content is located only one hop away from the consumer. The results of the experiment indicate that the built-in mechanism takes more hops due to its flooding mechanism, although it is only a single hop away. 

In Scenario-1-mid, the producer is represented by the green node, and the content is located at a moderate distance, two hops away from the consumer. The results revealed that the built-in mechanism required two additional hops compared to the proposed scheme, which had a difference of 2 hops compared to Scenario-1-near. Scenario-1-long is the producer in grey. This scenario involves the producer located farthest away, represented by the grey node. The results of this experiment revealed that the built-in mechanism takes significantly more hops than the proposed scheme to locate the content.


Figure \ref{fig:sim1-chart} shows the number of hops that the search experiences for the various considered distances (long, medium, or short distances, measured in number of hops). The results show that the conventional mechanism experiences a larger number of hops for content search than the proposed mechanism because the location of the content is not considered.

\begin{figure}[htbp]
    \centering
    \includegraphics[width =0.99 \columnwidth]{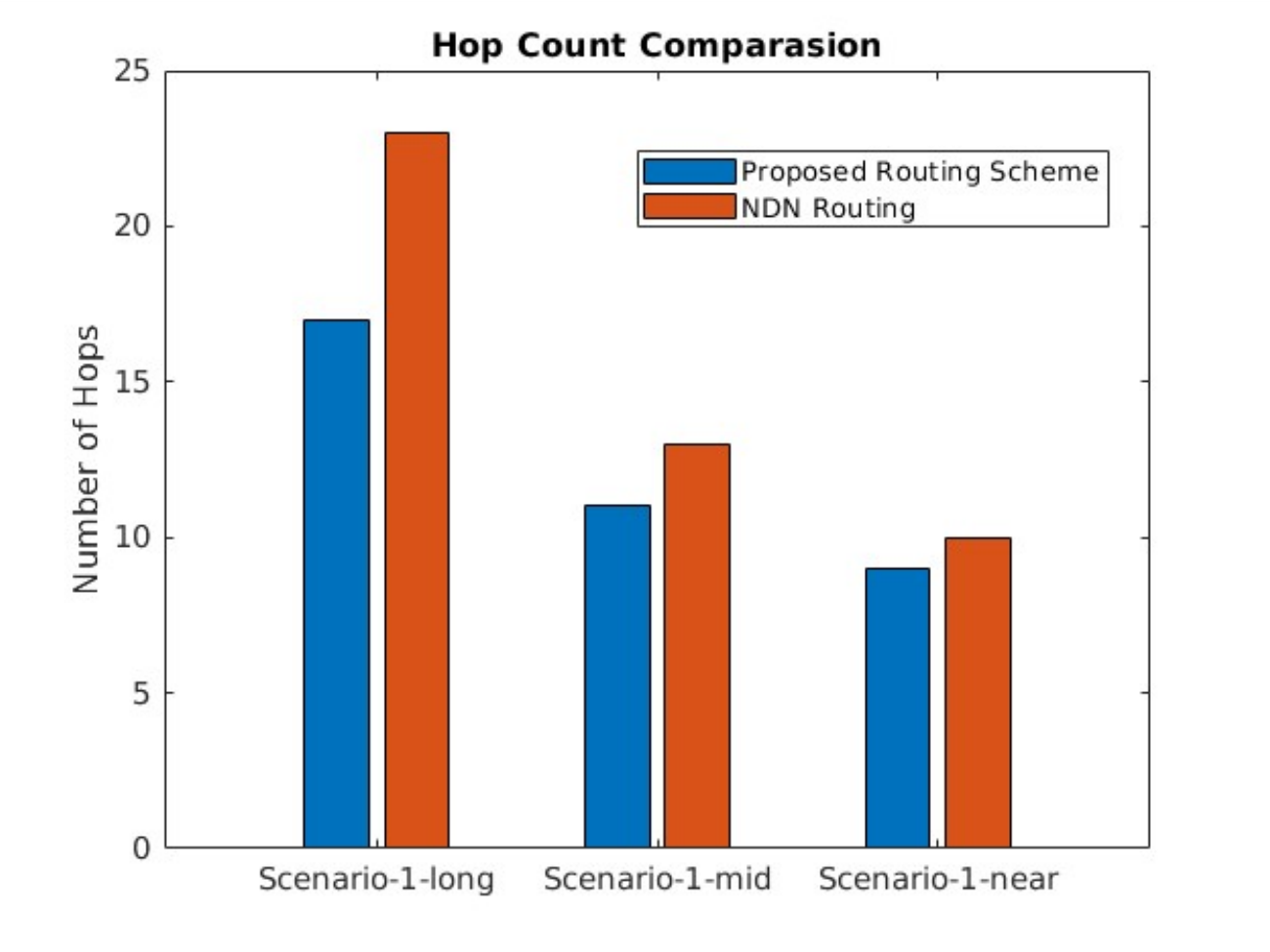}
    \caption{Results of the search mechanism of both the conventional method and the proposed one.}
    \label{fig:sim1-network}
\end{figure}

\subsection{Scenario 2: Populated NSFnet}

In this scenario, more hosts were added to the NSFnet network, resulting in a total of 55 consumers or nodes, as opposed to the 54 nodes in Scenario 1. In this experiment, one million unique data items were distributed across the eight NDN resolvers using a hashing mechanism, with 125,000 unique content items per resolver. Each consumer node requests unique content that is available in nodes in other networks. We measure  the total number of hops required to satisfy the requests. 
Table \ref{table:NN_DIFF} shows the results. 

\begin{table}
\label{table:NN_DIFF}
\begin{tabular}{ |p{1.5cm}||p{1.8cm}|p{1.8cm}|p{1.8cm}|  }
 \hline
 \multicolumn{4}{|c|}{Number of hops} \\
 \hline
 Cases & Distance (No. of hops) & Compared & Proposed \\
 \hline \hline
 (i)  & 1  & 10 &  8 \\ \hline
(ii) &  2  & 13  & 10 \\ \hline
 (iii) & 3 & 13 &  12\\ \hline
 (iv) & 4 & 28 &  14 \\ 
 \hline
\end{tabular}
\caption{Results of interest search in Scenario 2.}
\end{table}

\begin{figure}[htbp]
    \centering
    \includegraphics[width =0.99 \columnwidth ]{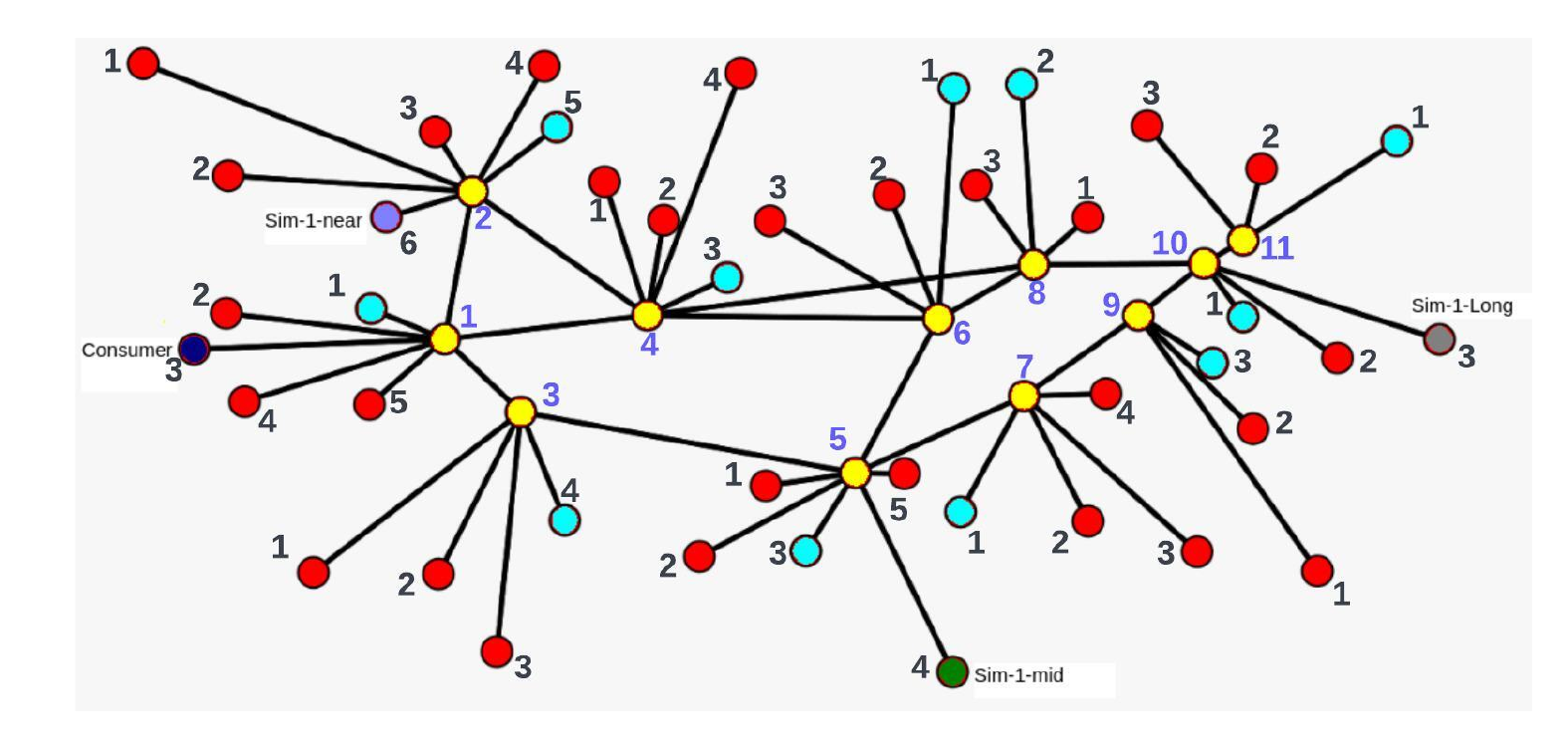}
    \caption{NSFnet-like topology used for evaluation. The red nodes are hosts, yellow ones are router nodes, blue nodes are NDN resolver clusters, and purple nodes are the ones selected as NDN resolver location for different distances.}
    \label{fig:sim1-network}
\end{figure}

\begin{figure}[htbp]
    \centering
    \includegraphics[width =0.99 \columnwidth]{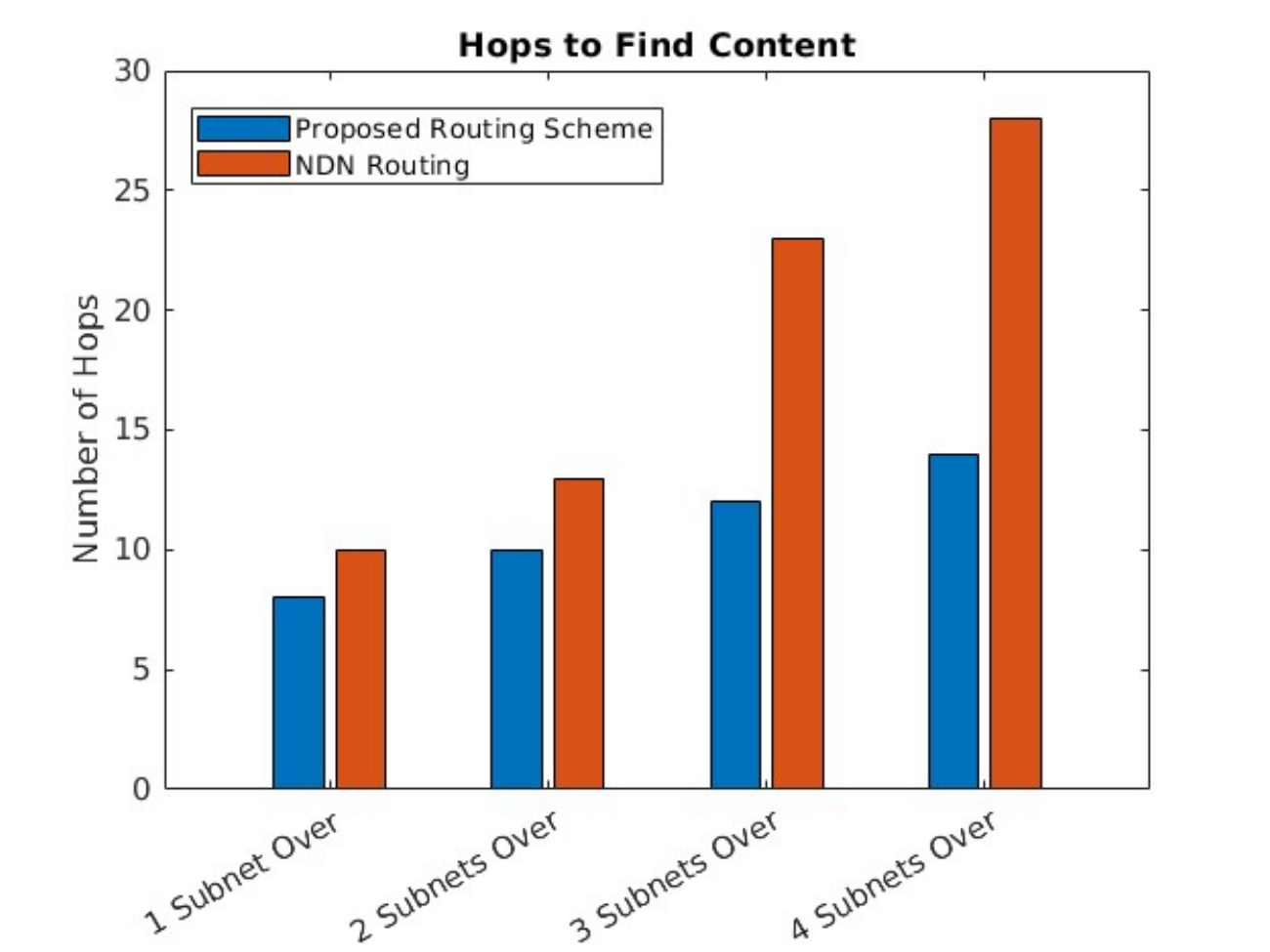}
    \caption{Number of hops taken by the two compared routing mechanisms to resolve a request.}
    \label{fig:simluation2}
\end{figure}

The results indicate that the proposed mechanism resorts to a smaller the number of hops find the content. 
The proposed mechanism is more targeted and efficient in finding the content,resulting in fewer hops. A graphical representation of the results can be seen in Figure \ref{fig:simluation2}.

\section{Scenario 3: Scalability in a Large Network}
\begin{figure}[htbp]
    \centering
    \includegraphics[width =0.99 \columnwidth]{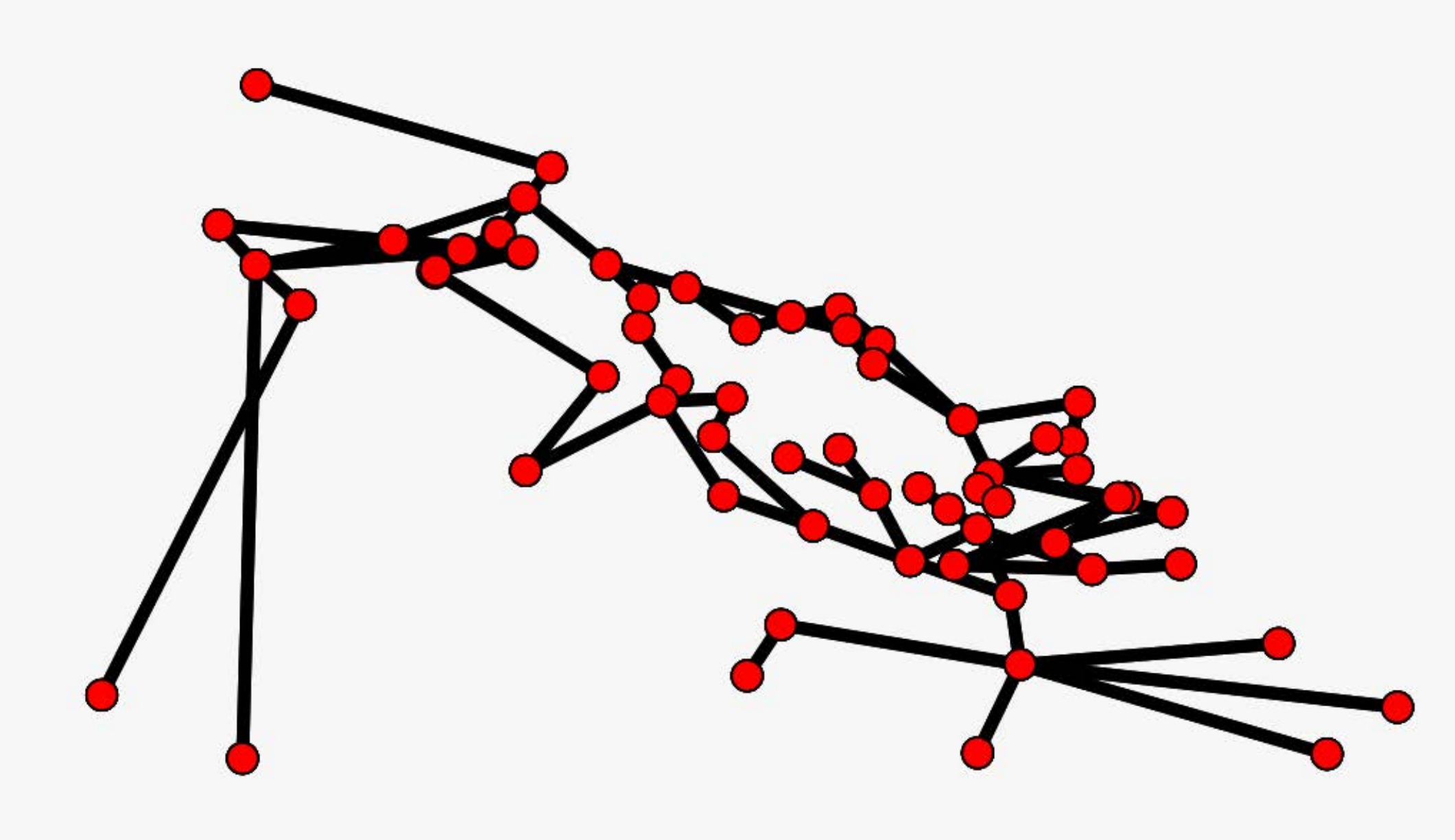}
    \caption{OTEGlobe architecture used for extended distance tests.}
    \label{fig:OTEGlobe}
\end{figure}

To evaluate the scalability and robustness of the proposed scheme, a larger topology was selected: the OTEGlobe network. 
Figure \ref{fig:OTEGlobe} illustrates this topology used for testing scalability and to stress test the proposed scheme. The topology consists of 427 nodes, including 305 hosts, 61 routers or root nodes, and 61 NDN clusters. 

\begin{table}
\label{table:scenario3}
\begin{tabular}{ |p{1.5cm}||p{1.8cm}|p{1.8cm}|p{1.8cm}|  }
 \hline
 \multicolumn{4}{|c|}{Number of hops} \\
 \hline
 Cases & Distance (No. of hops) & Compared & Proposed \\
 \hline \hline
1  & 9  & 10 &  8 \\ \hline
2 &  14 & 13 & 10 \\ \hline
3 & 19 & 13 & 12\\ \hline
4 & 23 & 28 & 14 \\ \hline
5 & 28 & 10 & 16 \\ \hline
6 & 33 & 13 & 18 \\ \hline
7 & 38 & 13 & 20\\ \hline
8 & 42 & 28 & 22 \\ \hline
9 & 47  & 10 & 24 \\ \hline
10 & 51  & 13 & 26 \\ \hline
11 & 57 & 13 & 28\\ \hline
12 & 62 & 28 & 30 \\ \hline
13 & 68  & 10 & 32 \\ \hline
14 & 73 & 13 & 34 \\ \hline
15 & 78 & 13 & 36 \\ \hline
16 & 83 & 28 & 38 \\ \hline
\end{tabular}
\caption{Results of Scenario 3 test: number of top 5\% hops.}
\end{table}

\begin{figure}[htbp]
    \centering
    \includegraphics[width =0.99 \columnwidth]{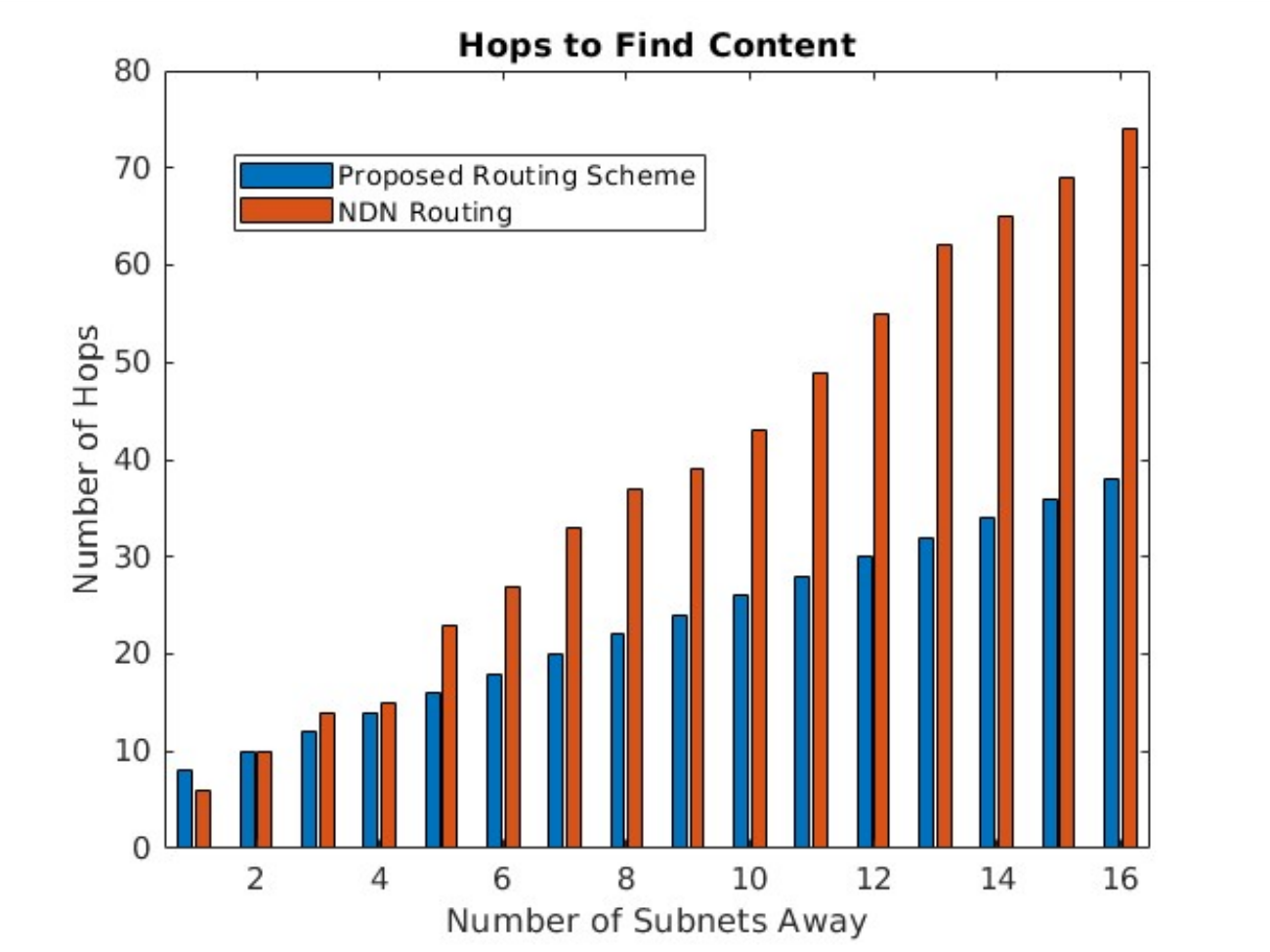}
    \caption{Comparison of searched number of hops for difference distances.}
    \label{fig:sim-3}
\end{figure}

As shown in Figure \ref{fig:sim-3}, the results indicate a clear contrast between the built-in and the proposed schemes, in terms of scalability and resource utilization.

\begin{figure}[htbp]
    \centering
    \includegraphics[width =0.99 \columnwidth]{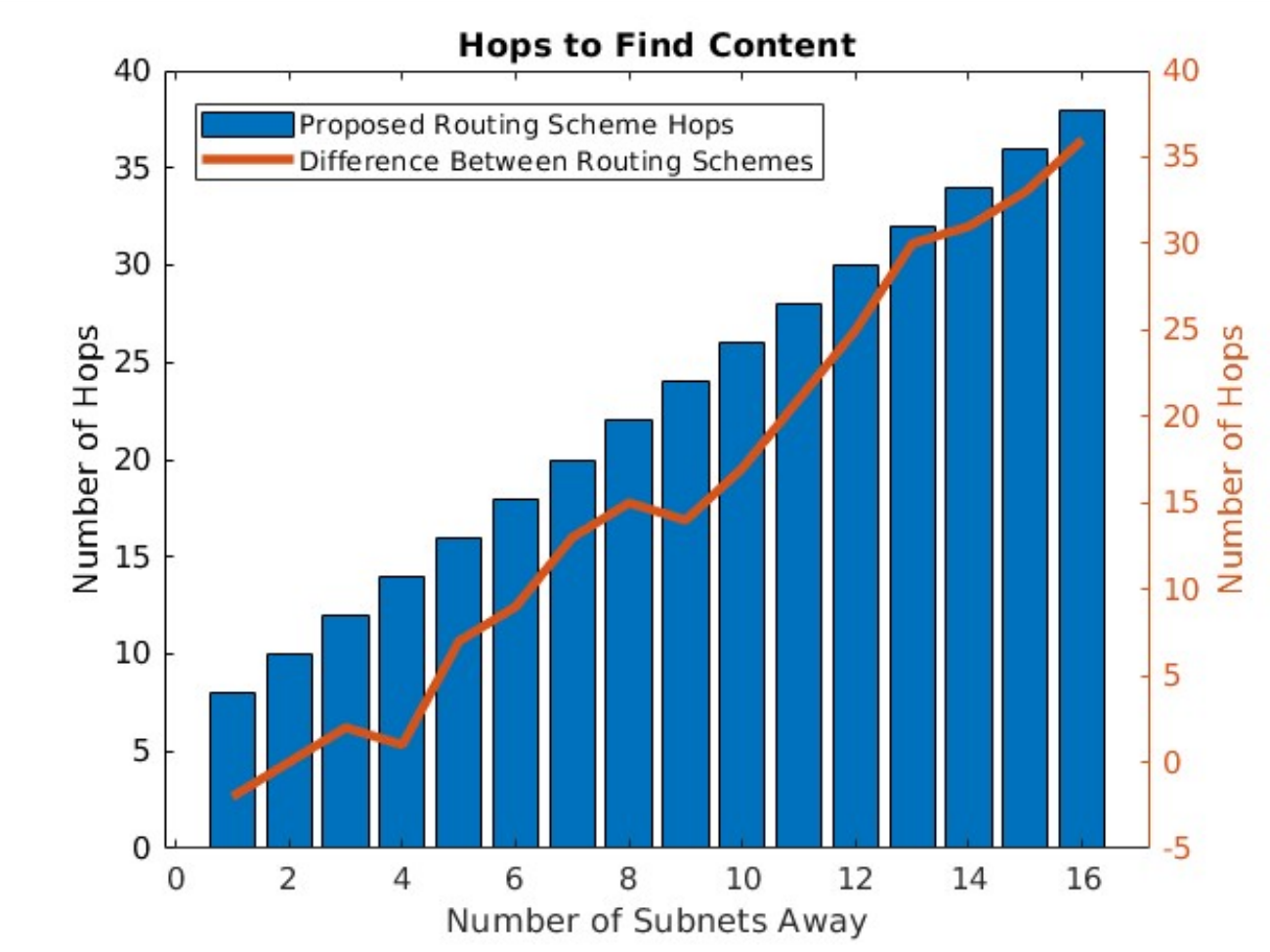}
    \caption{The bar graph shows exponential increase in the number of hops as the distance to the node with the content increases. The line indicates the difference between the compared and proposed approach.}
    \label{fig:sim-3-line}
\end{figure}

Figure \ref{fig:sim-3-line} shows the difference between the routing schemes by the line plot and also the number of hops taken by the proposed method. These results show that the proposed scheme almost halves the number of hops taken by the conventional scheme. At the same time, this results shows that the proposed mechanism is scalable, whereas conventional NDN routing presents challenges.

\subsection{Scenario 4: Nonuniform Distribution of Content}

In this scenario, we tested the efficiency and performance of the hashing algorithm when the content is not distributed uniformly. 
This test was done twice; the first one with three NDN resolvers, and then with eight NDN resolvers. Each experiment was performed 10 times and then averaged.

In shown in Table \ref{table:scenario4-1}, the total content amount of unique content distributed in the network is 400,000 objects. This content is divided into 200,000 objects for NDN1, and the remaining content is equally distributed between NDN2 and NDN3. The results show that the time taken to find content in NDN1, which has more content, exhibits a marginal difference in microseconds. Similar results are shown in Table \ref{table:scenario4-1}, where the content was distributed even more unevenly. NDN1 has 300,000 objects and the remaining 100,000 are split among NDN2 and NDN3. These results showed that the time difference is negligible. Even more, the time taken for NDN1 is less than in the table. This shows that a difference of 200,000 or 100,000 is not sufficient to degrade the performance of the hashing algorithm.
\begin{table}
\caption{Different content distribution on three networks.}
\label{table:scenario4-1}
\begin{tabular}{ |p{1.3cm}||p{1.3cm}|p{1.3cm}|p{1.3cm}|p{1.3cm}|  }
 \hline
 \multicolumn{4}{|c|}{Number of hops} \\
 \hline
 Network & Content (No. items) & Time & Content (No. items) & Time \\
 \hline \hline
 NDN1  & 200,000  & 0.12676 &  300,000 & 0.12288\\ \hline
NDN2 &  100,000  & 0.12073  & 50,000 & 0.12132 \\ \hline
NDN3 & 100,000 & 0.12218 &  50,000 & 0.12218 \\ \hline 
\end{tabular}

\end{table}

In Table \ref{table:scenario4-2}, the same experiment was performed, but at a larger scale. We use 1,000,000 content pieces and eight NDN resolvers. This test showed that NDN1 takes approximately 0.3 ms. This means that with 500,000 content, the hashing algorithm would experience a significant increase in execution time. But, an increase of 0.3 ms is almost unnoticeable to the end user. So, the hashing data structures can sustain increase in content loads.
\begin{table}
\caption{Different content distribution on eight networks/resolvers.}
\label{table:scenario4-2}
\begin{tabular}{ |p{1.3cm}||p{1.3cm}|p{1.3cm}|p{1.3cm}|p{1.3cm}|  }
 \hline
 \multicolumn{4}{|c|}{Number of hops} \\
 \hline
Network & Content (No. items) & Time & Content (No. items) & Time \\
 \hline \hline
NDN1  & 125,000  & 0.12676 &  650,000 & 0.14334 \\ \hline
NDN2 &  125,000 & 0.12073 & 50,000 & 0.11773 \\ \hline
NDN3 & 125,000 & 0.12218 & 50,000 & 0.113 \\ \hline 
NDN4  & 125,000  & 0.12676 & 50,000 & 0.12676\\ \hline
NDN5 &  125,000  & 0.12073  & 50,000 & 0.10693 \\ \hline
NDN6 & 125,000 & 0.12218 &  50,000 & 0.11478 \\ \hline 
NDN7  & 125,000  & 0.12676 & 50,000 & 0.1137 \\ \hline
NDN8 &  125,000  & 0.12073  & 50,000 & 0.10934 \\ \hline
\end{tabular}

\end{table}

\section{Conclusions}
\label{sec:conclusions}

The conventional NDN content search mechanism (here called routing) uses flooding, which can lead to network congestion and resource exhaustion. To address this issue, we proposed a distributed nameserver lookup service, similar to DNS, with a load-balanced hashing mechanism for efficient distribution of content. The proposed scheme was simulated, and the results show that BalanceDN finds content faster than the conventional search method used in NDN. The proposed scheme efficiently distributes and locates the name of the content on the distributed nameserver lookup service nodes, thereby eliminating the need for flooding the network with Interest packets. The results also show that the proposed scheme scales linearly and uses far fewer resources as compared to the built-in mechanism.
Overall, the proposed routing scheme offers an efficient solution for routing and resolving data in NDN. By reducing network congestion and resource usage, it has the potential to improve the performance and reliability of NDN-based applications and services. Furthermore, the proposed scheme can be further optimized and extended for large-scale and real-world deployment, making NDN a feasible option for future Internet architecture.

\end{document}